# Unique Device Identification Based Linkage of Hierarchically Accessible Data Domains in Prospective Hospital Data Ecosystems

Karol Kozak[1], André Seidel[2], Nataliia Matvieieva[2], Constanze Neupetsch[2,5], Uwe Teicher[2,*], Gordon Lemme[2], Anas Ben Achour[2], Martin Barth[4], Steffen Ihlenfeldt[2,3], Welf-Guntram Drossel[2,5]

[1] Technische Universität Dresden, Center for Evidence-Based Healthcare, 01307 Dresden, Germany
[2] Fraunhofer Institute for Machine Tools and Forming Technology IWU, 01187 Dresden, Germany
[3] Technische Universität Dresden, Chair of Machine Tools Development and Adaptive Controls, 01069 Dresden, Germany
[4] Fraunhofer Institute for Ceramic Technologies and Systems, Maria-Reiche-Str. 2, 01069 Dresden, Germany
[5] Technische Universität Chemnitz, Professorship for Adaptronics and Lightweight Design in Production, D-09107 Chemnitz, Germany

* Correspondence: uwe.teicher@iwu.fraunhofer.de; Tel.: +49-351-463-34385

**Abstract:** The electronic health record (EHR) targets the systematized collection of patient-specific electronically-stored health data. Currently the EHR is an evolving concept driven by ongoing technical developments and open or unclear legal issues concerning used medical technologies, data integration from other domains and unclear access roles. This paper addresses cross-domain data integration, data fusion and access control using the specific example of a Unique Device Identification (UDI) expanded hip implant. In fact, the integration of technical focus data into the hospital information system (HIS) is discussed and presented based on surgically relevant information. Moreover, the acquisition of social focus databased on mHealth is approached, which also covers data integration and networking with therapeutic intervention or acute diagnostics data. Data integration from heterogeneous domains is covered while using a data ecosystem with hierarchical access based on a shell embedded role model, which includes staggered access scenarios.

## 1. Introduction

Unique Device Identification (UDI) is a system used to identify devices within the healthcare supply chain based on a consistent, standardized, and unambiguous machine-readable identifier to keep track on the post marketing performance of medical devices [1]. The performance of e.g. an hip implant cannot be evaluated without considering individual factors of the recipient [2] and the conditions of therapeutic intervention [3]. Consequently, patient, medicine and product has to be linked and monitored to enable a well-founded evaluation. Regardless the still missing legal framework conditions [4] and the existing ethical and political questions [5], digitalization of the health system is advancing [6] which is beneficial towards a holistic assessment. Either way, part of this development is the cross-domain linkage of data [7], that at least can be resolved on a patient-by-patient basis to prepare for intelligent data analysis. This requires beyond analysis objectives an appropriate data infrastructure, which enables different data domains to be linked while providing adequate hierarchical data access concepts. Consequently, this paper approaches the framework for cross-domain cooperation and intelligent data analysis in a specific application scenario embedded in prospective digital hospital ecosystems.

## 2. Framework conditions for digitization in individualized medicine

### 2.1. Intelligent data analysis framework

An electronic health record (EHR) is the systematized collection of patient and population electronically-stored health data in a digital format. These records can be shared across different health care settings. Records are provided through network-connected, enterprise-wide information systems or other information networks and exchanges. EHRs can include a range of data such as individual risk assessments, health monitoring, acute diagnostics and therapeutic intervention while enabling altogether intelligent data analysis which, according to Hahn et al. [8], is termed as information cycle of personalized medicine (Fig. 1).

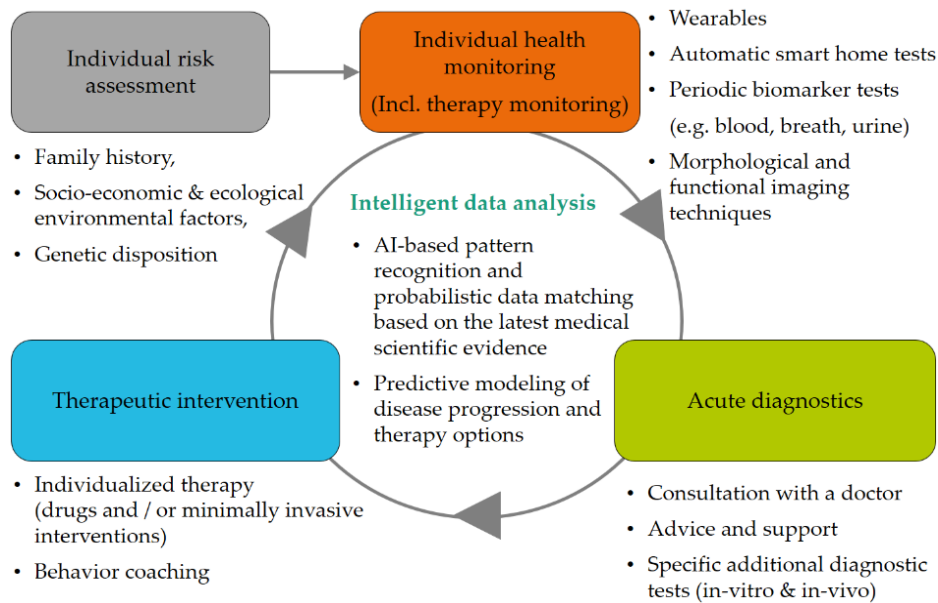

**Figure 1.** Information cycle of individualized medicine [8].

This data is stored in centralized Hospital Information System (HIS) in form of text, images (DICOM and other type), scans as patient digital twin. EHR enable patients and hospital to manage their health information in public (e.g. hospital) and private environments as Personal Health Record (PHR). The information contained in EHR is highly sensitive. Unintended exposure of this data threatens an intimate part of a patient's private sphere and may lead to undesirable consequences. EHR is a communication tool that supports clinical decision-making, coordination of services (sick type, care type), evaluation of the quality and efficacy of care, research, legal protection, education, and accreditation and regulatory processes. It is the business record of the health care system, documented in the normal course of its activities. Patients routinely review EHR and are keeping PHR in own digital archive or in patient portals (e.g. at health insurance company such as "TK-Safe", "Vivy" or "AOK-Gesundheitsnetzwerk" in Germany) while the patient is the owner of EHR and PHR. The physician, practice, or organization is the owner of the physical medical record because it is its business record and property, and the patient owns the information in the medical record. Although the record belongs to the care facility or doctor, it is truly and also the patient's information. EHR should be released to other stakeholders only with the patient's permission or as allowed by law or studies: public registry, ministries. PHRs are already on the market. Purpose of PHR is to maintain good health and target outcome, this could include daily vital signs (blood pressure, heart rate, etc.), number of walking steps, amount of exercise, and calorie intake. In addition to these, information for medical use might be considered, such as blood type, allergies, pre-existing diseases, medicine the user is taking, emergency contact information, and information about the user's medical institution.

*2.2. Synergy potential in information linkage using the example of a hip implant*

Neugebauer [9] states that the digital transformation is promoted by the interaction of technologies which were previously perceived as independent of each other. Hahn et al. [8] note in this context, that networking of the individual sectors and the structured use of the integrated information are still pending. Moreover, Hahn et al. [8] conclude that sustainable success can only be expected if the interlinking of technological and biomedical research on the one hand and clinical implementation and product development on the other is permanently guaranteed. Either way, this paper approaches data-driven interdisciplinary research from an application-oriented perspective in an incident-based scenario (Figure 2). This example illustrates selected dependencies between social parameters, medical factors and technical aspects being important for surgery and healing which are currently not linked sufficiently. In fact, it is a common practice to obtain this information laborious on a case-by-case basis, which requires an appropriate lead-time before the operation, which is a major disadvantages in the case of emergency medical treatment. Consequently, this paper addresses this shortcoming and develops a possible solution scenario showing how this information can be linked at the EHR level. Moreover, it is shown how this information can be made accessible based on implant inherent features while introducing a role model for access regulation and data protection.

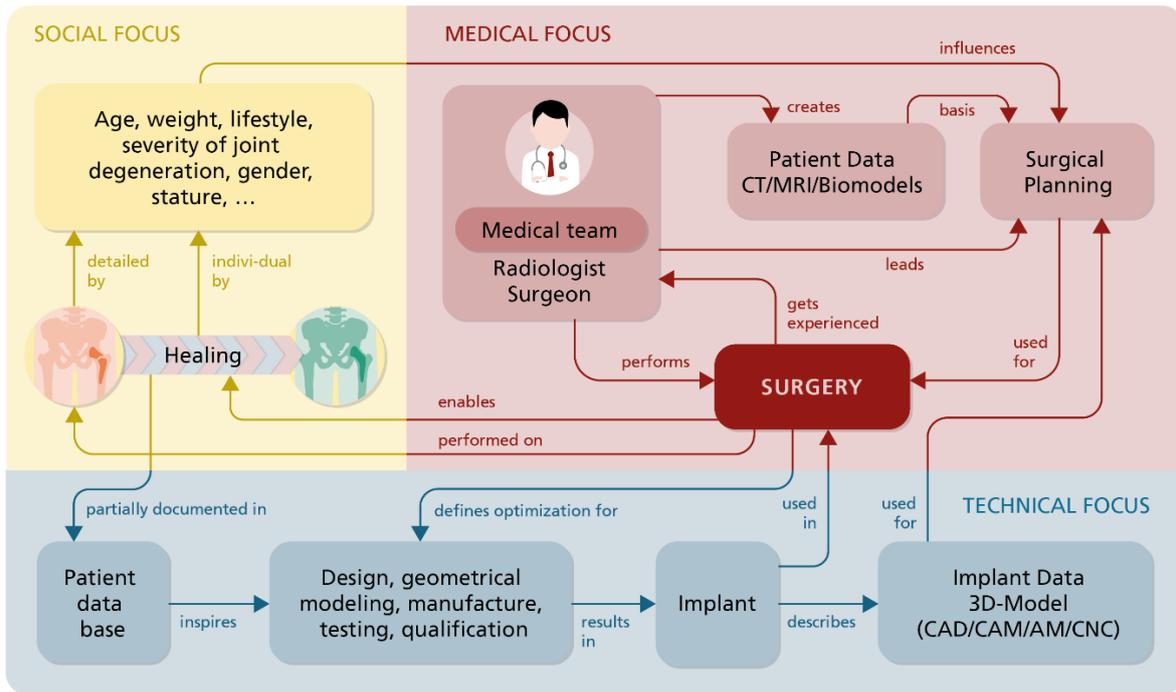

**Figure 2.** Incident based networking of social aspects, medical factors and technical aspects.

## 3. Data-driven networking of information

### 3.1. Regulatory demands

In the medical technology industry, quality assurance is a particular focus due to the stringent regulation. With a changeover period, the European Medical Device Regulation (MDR) EU 2017/745 comes into effect in May 2021 and replaces the European directives on medical devices. In the future, the regulation will obligate manufacturers to mark medical devices that are marketed in the European Union with unique codes. The main objective for the introduction of these codes is to increase patient safety. Unique product identification prevents confusion of medical devices and makes counterfeiting more difficult. The markings are implemented by the UDI system. The UDI enables to track medical devices e.g. from the last step of post-processing in manufacturing, where the marking is applied to the component (e.g. by laser engraving). Nevertheless, the medical products can only be tracked from this point through the logistics process to the hospital. Figure 3 illustrates selected phases of a medical device life cycle covering design, manufacturing, post-processing, logistics and the union with the patient. Unfortunately, a doubtless identification of an implant after implantation is impossible with the UDI system, which largely excludes proof of quality and originality after implantation. Across industries, product piracy is a major problem and the estimated economic damage has been increasing over the years [36]. In addition to the generally valid comments on product piracy and its consequences, other aspects have to be considered in the field of medical technology. Since the publication of the "Implant Files" in 2018 by a group of investigative journalists, the problem of defective medical devices became a sociopolitical issue [37].

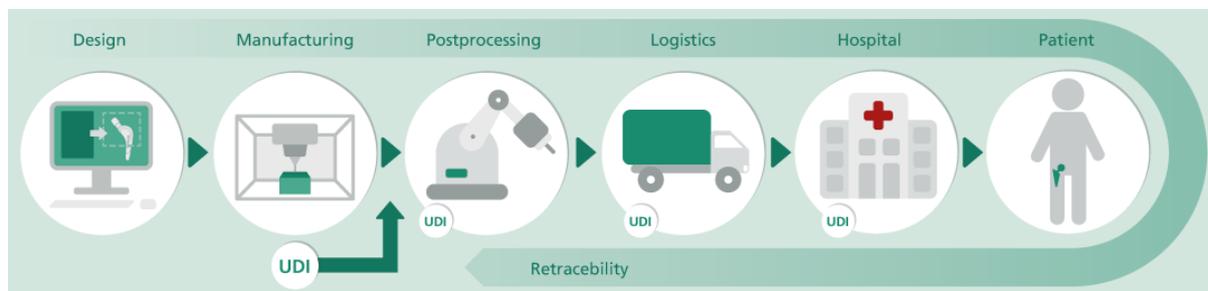

**Figure 3.** Retraceability of a medical device with a UDI.

### 3.2. Component-inherent identifier based data access

Inherent markings of implants could be a solution to enable counterfeit-proof while satisfying the regulatory requirements for machine-readable marking in the form of a barcode or data matrix. In addition, traceability can be expanded back to the design process where the inherent markings are integrated. Moreover, this enables simultaneously to also include the manufacturing process into the traceability chain as well (c.f. Fig. 3). Either way, the greatest potential for innovation is seen in the ability to clearly identify the implant after surgery. The latter, preferably via the usage of non-invasive technologies already available in the hospital, or technologies, which can be provided without great technological effort and financial investment. Either way, the inherent feature could act as a key to access distributed information (c.f. Fig. 2) after signal processing and appropriate decoding (Fig. 4).

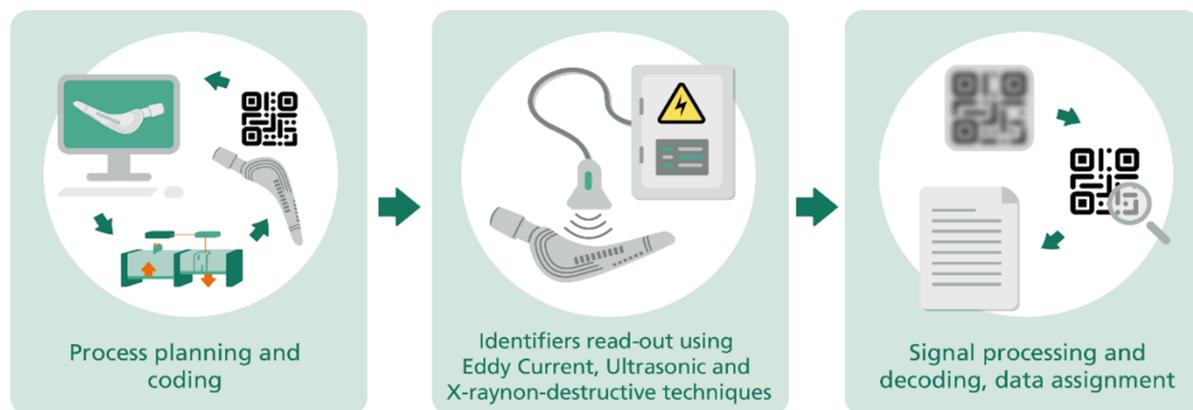

**Figure 4.** Principle of inherent markings in an additively manufactured implant with part identification and tracking.

Matvieieva et al. [10] performed identifier read-out applying Eddy Current (EC), Ultrasonic (US) and Micro-CT (also in X-ray mode) as non-destructive methods. The EC, US and X-Ray techniques allow receiving a part inherent data encoded in 1D- and 2D-codes. The feasibility of the methods was shown for 1D-Pharmacode, UDI 1D-Barcode (ISO 128) and UDI 2D-Datamatrix for Titan, Titan Alloy and Stainless Steel. Due to the physical restrictions of the chosen non-destructive methods (e.g. penetration depth, magnetic and acoustic waves transmission, propagation and density, etc.), obtained identifiers' signals have to be processed. In fact, after the identifier´s read-out the obtained signals are processed by morphological and mathematical operations and being decoded, enable to be further linked with data or information stored in a database (Fig. 4).

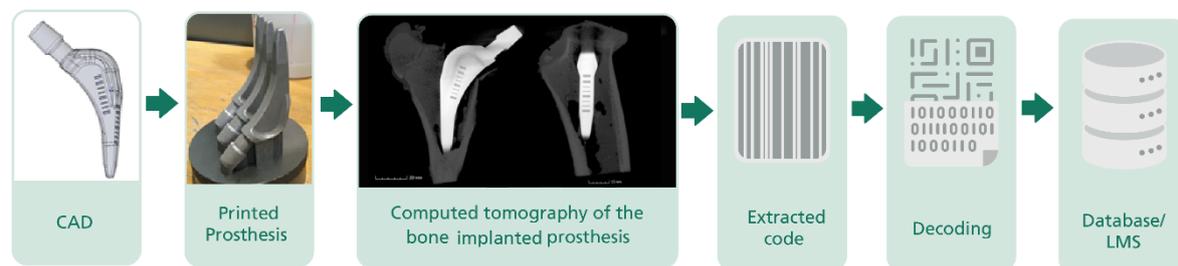

**Figure 5.** Example of the components (parts) identification on example of integrated Pharmacode into the hip prosthesis and identifier's data extraction in implanted state by computed tomography.

### 3.3. Technical focus data

Starting point is a construction of a 3D model of hip implant (Fig. 5), based on anthropometric data or even patient-specific information in which the coding is integrated as well. Considering typical hip implant sizes, geometric complexities and/or quantities, 3D printing (additive manufacturing) is increasingly developing as a competitive manufacturing method [11]. Suitable Additive Manufacturing (AM) procedures are primary selective laser melting (SLM) [12, 13] or electron beam melting (EBM) [14, 15]. In addition, the layer-wise build-up and the achievable resolution are very beneficial for the integration of inherent features during manufacture (c.f. Fig. 3). Using these manufacturing processes, the 3D data model needs to be converted into a facet model first (mesh) [16]. In the following, the parts are positioned in the build-chamber, supported and sliced using standard AM software [17]. Based on material selection criteria such as biocompatibility, Young´s modulus, strength and fatigue strength a (certified) raw material is selected [18–20], which is particularly tailored to the AM process requirements [21]. Material selection criteria like particle size and particle size distribution, morphology or chemical properties are continuously checked and monitored [22, 23]. The e.g. SLM process data has to be qualified for applications like implants and is subdivided in predefined parameters (considered static) and parameters to be controlled by in-process sensing (considered dynamic) [24]. Either way, essential parameters are monitored and archived [25–27]. The same applies to process data from post-heat and/or pressure treatment [28] as well as from destructive (DT) and non-destructive material testing (NDT) [29]. Moreover, mechanical post-machining, performed to adapt the implant to the recipient needs, generates data [30]. An example of this is the description of the geometric interface to the patient (bone–implant interface) including parameters like e.g. surface roughness. The result, however, is an extensive description of the implant; the implant creation process as well as additional information such as corresponding implant tools (Fig. 6). Obviously, parts of this information are of interest to surgeons (medical focus), whereas information about the implant recipient and the use of the implant (social focus) are of interest to the manufacturer (Fig. 2). This means, regardless of the individual legal framework, the possibility of a controlled linkage of information is seen as desirable.

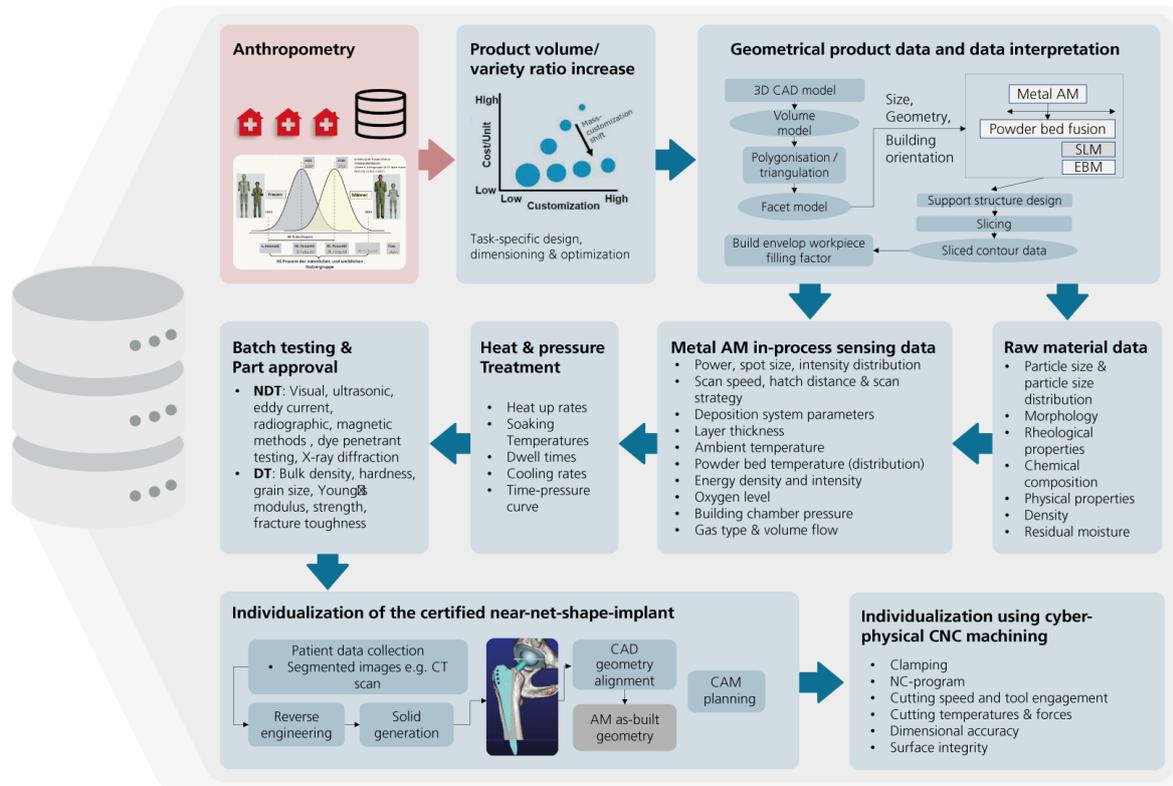

**Figure 6.** Schematic illustration of the process chain for additively manufacture and subtractive individualization of hip implants with a selection of quality-determining parameters monitored and archived in process databases as part of quality assurance.

*3.4. Social focus data*

The social background of an implant-receiving patient (Fig. 2), includes individual characteristics and conditions such as age, gender, lifestyle and constitution type, which is of crucial importance for the formation of diseases, their duration and treatment. The continuous monitoring or even documentation of this background can be ensured by a variety of technologies from the field of mHealth, which allow a broad mapping of dynamic data sets such as lifestyles and physical activities [31]. MHealth is an aspect of eHealth although there is not a universal definition of mHealth [1]. However, there is consensus that mHealth can be understood as medical and public health practice supported by mobile devices, such as mobile phones, patient monitoring devices, personal digital assistants (PDAs) and other wireless devices [32]. This means that mHealth can be understood as "the use of mobile communications for health information and services" as patient-individual behavior without direct involvement of the health service provider [33]. Here, mHealth is seen as the technical prerequisites to monitor and document the social focus data during healing (Fig. 7) or even beyond (c.f. Fig. 2).

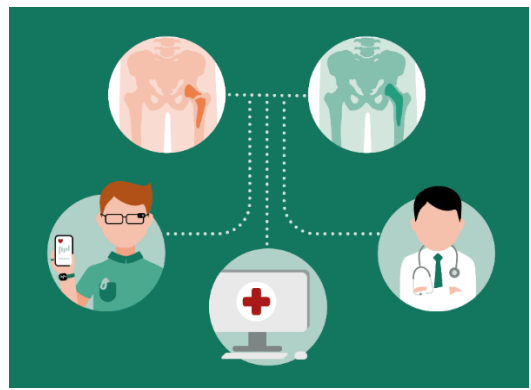

**Figure 7.** Illustration of close monitoring of healing using mHealth applications.

Besides the fact that mobile communication and audio-visual interaction is the decisive enabler for mHealth, there is also the aspect of powerful sensor-based hardware and high flexibility in software development for smartphones and, to a limited extent, wearables [34]. Widespread sensor systems for smartphones are [35, 36]:

- Light sensor technology (ambient light, camera system in combination with lighting),
- Proximity sensors,
- Acceleration sensor technology,
- Rotation sensor technology (gyroscope),
- Electromagnetic sensor technology,
- Digital compass (magnetometer),
- Acoustic sensor technology via microphone,
- Sensor technology for location tracking (Global Positioning System GPS).

In contrast to smartphones, wearables are more specific for a particular application and therefore have higher specificity for the integrated sensors, so that headbands can, for example, derive targeted electroencephalography (EEG) [37]. This means that a targeted data collection is possible with the help of sensor technology, internal processing by specific software (e.g. apps), visualization based on this data and a mostly wireless communication to third parties in form of a uniform data image. For example, the following parameters can be acquired by means of wearables and mobile devices:

- Heart rate and pulse oxymetry readings with photoplethysmography [38],
- Systems to monitor activity and sleep [39].

Hence, mHealth is seen as an enabler that contributes to rehabilitation by providing vulnerable data about the rehabilitation measures and patient-specific activity (Fig 6) which can be stored in the EHR. In addition, realistic load scenarios can be determined which could contribute to the further optimization of hip implants (Fig. 2). This in order to promote healing [40] while e.g. targeting shorter hospital stays and lower treatment costs [41]. In addition, mHealth can help to achieve a consistent data base (Fig. 2) being used to evaluate the optimal intensity, frequency and effects of rehabilitation from a wide variety of patients over a longer period of time as this data is currently not available or insufficient [42]. Either way, linking social focus data and production data could enable significant improvements with regard to e.g. determine the actual wear of the implant based on e.g. posture, weight and movement profiles summarized here as social focus data (Fig. 2).

*3.5. Medical focus data*

Both routine diagnostics and revision surgery require information about the implant were placed decades ago. A specific identification using only medical imaging is currently not possible and requires access to the documentation of the initial treating physician or hospital (Fig. 2). For elective as well as for acute medical interventions, this documentation is not available or only available with enormous effort. With inherent markings in the implant it is feasible to obtain information relevant for revision surgery or routine diagnostics even after the insertion of an implant into the human body (Fig. 4). To assure a valid and reliable surgical planning, medical data of past treatments especially the surgical processes, follow-up examinations and rehabilitation measures are needed. Insufficient information about the implant and medical focus data (Fig. 2) could lead to complications during revision surgery and therefore considerable efforts are being made to obtain this information, which significantly extends the preoperative time in the hospital causing additional costs. In fact, an example for necessary information about the particular implant concern the appropriate revision instruments and the existing implant components [43–45]. Moreover, the research effort prior the operation is continuously increasing which is linked to the increasing number of operations and the growing variety of implant types, sizes, variants, and material combinations. In addition to the knowledge on the implant system used, the information on the initial implantation process, which cannot be seen in the medical images (Computer Tomograph, X-Ray), are highly relevant for a gentle and successful revision of the hip implant. In fact, insufficient information clearly increases the risk of complications for the patient. Here it shall be emphasized, that the use of unsuitable revision instruments causes the hazard of an enlargement of the wound surface through an invasive procedure. This, in turn, increases the risk of infection and bleeding or even periprosthetic fracture. In addition, the inevitable prolongation of surgery time can lead to a higher anesthetic risk, an increased risk of thrombosis and unstable cardiovascular function. Consequently, there is the obligation to report "incidents" in connection with medical devices prescribed by the German Federal Institute for Drugs and Medical Devices (BfArM), which includes an indication of the cause. Hence, unique inherent identification (Fig. 4) and (partial) networking of information (Fig. 2) in an transparent and retraceable database seems promising to assure a higher quality of healthcare [39] if unauthorized access is avoided.

## 4. Database and data access

*4.1. Data integration from heterogeneous domains*

Data integration is a crucial issue in the environments of heterogeneous patient – production data sources (Fig. 2). First, there are heterogeneous data types and formats located in different databases, which implies to solve data integration challenges as a prerequisite for gaining useful information and knowledge based on appropriate analytical methods. The Laboratory Management System 4.0 (LMS4.0) [46] applied here was especially developed for this purpose. LMS 4.0 enables to request data from different locations (e.g. surgery, the HIS and/or an implant producer) as a routine using web user interfaces. Using LMS 4.0, the surgeon collects e.g. Magnetic Resonance Imaging (MRI) results from HIS, checks patient data stored there and has access to the integrated technical focus data (Fig. 8) while planning the operational approach.

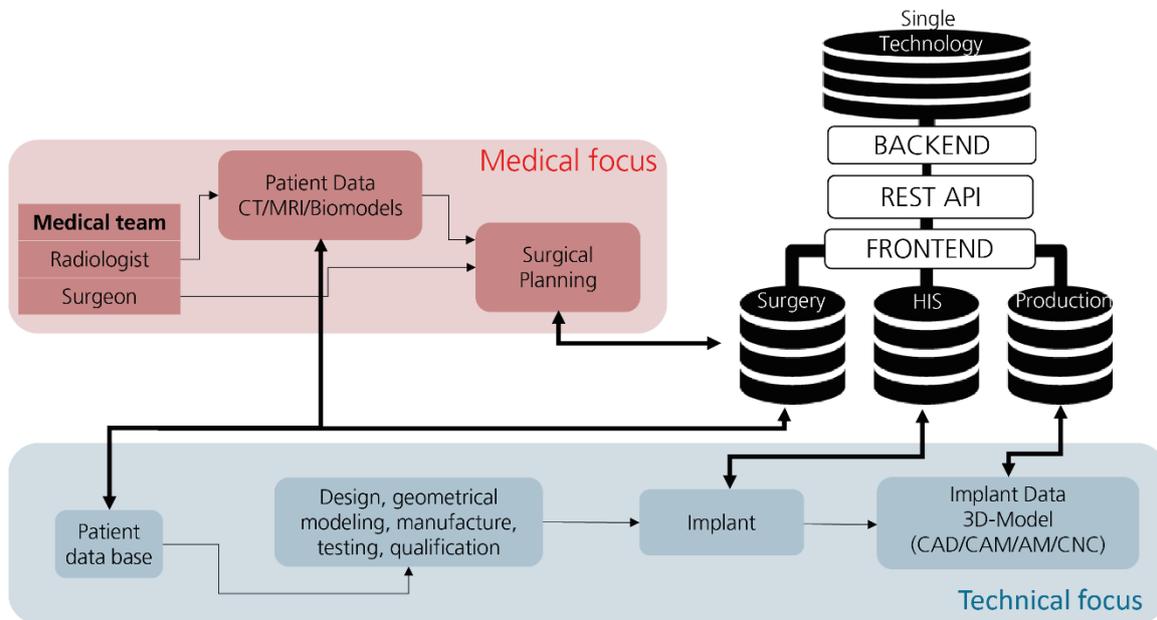

**Figure 8.** Partial networking of distributed data and information from different domains.

Based on integrated patient and production data, which can be expanded to an implant database that covers several manufacturers, surgical planning is simplified in order to e.g. determine which bone implant is the most suited in the particular case. Moreover, LMS 4.0 generates a dashboard and a report that helps the operating staff to prepare the surgery with the selected implants through the automatic output of the associated tools and instruments. Moreover, the surgeon can use this report to prove his preparations for the procedure and to comprehensively explain the operation to the patient. LMS4.0 presents an architecture which implements data integration in hospital from the production, surgery preparation and patient data. LMS 4.0 integrates databases without any changes to the individual databases (Structured Query Language (SQL) database, software backend, Application Programming Interfaces API, Frontend) nor any need to maintain another database. The solution combines database technology and a wrapper layer known from Extraction Transformation Loading systems and brings it to SQL Database, WEB API (backend) layer, Interface layer (Rest API) and frontend. It also provides semantic integration through connection mechanism between data elements. The solution allows for integration of patient, surgery, production data in one technological framework: data management platform and implementation of analytical methods in one end-user environment. The patient data (see Fig. 2) is transferred, secured, to a HIS. Medical data storage in LMS 4.0 offers a highly scalable clinic web storage service that uses cumulative digital objects (eg. patient, surgery, implant) rather than blocks or files. Object storage typically stores data, along with metadata that identifies and describes the content. For metadata management and automated quality control and data fusion (ETL Processes) a data consistency model (LMS I4.0 metamodel) is used to enable eventual consistency for updates or deletes to existing objects.

*4.2. Role concept for secure data access*

On the basis of the information domains and the links and interfaces shown in Fig. 2, it becomes obvious that the database structure, the IT system design in the back end (c.f. Fig. 8) has to accommodate different user roles to protect secure data access to sensitive patient data. Consequently, a role model was developed that takes into account both, different users or user groups (e.g., patients, medical staff, manufacturers of medical products, and first aid providers) as well as special situations (e.g. emergency access). Deviating from static access models (role-based access control) as well as traditional shell models in this case pure login information was linked with additional contextual information (attributed-based access control) in order to allow hierarchical access control. Either way, the principle is shown in Fig. 9 indicating the four basic roles embedded in the defined shell model.

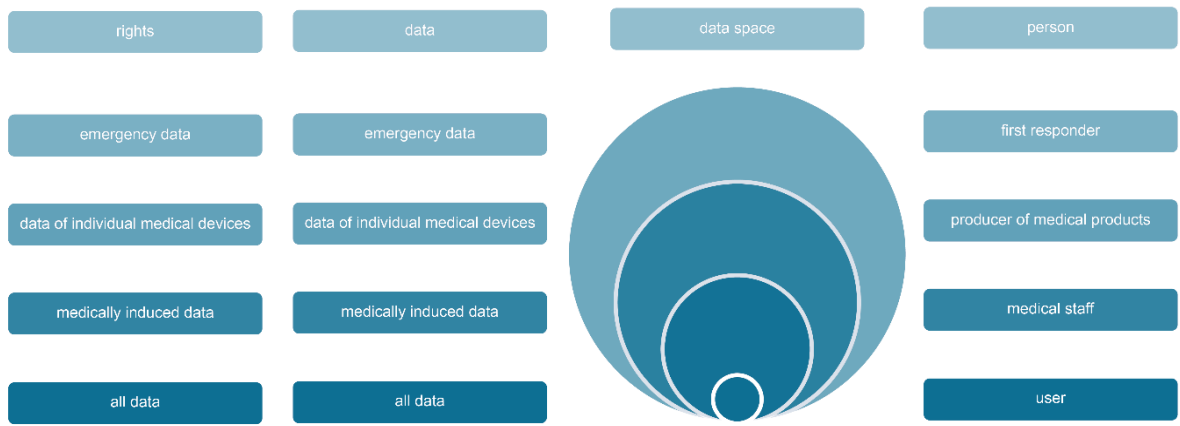

**Figure 9.** Principle of hierarchical data access based on a shell embedded role model.

Nevertheless, the person category is subdivided by four roles: "first responder", "producer of medical products", "medical staff" and "user". Users can view all data via a terminal device after registration, e.g. with a digital health card equipped with a radio-frequency identification (RFID) transponder (radio-frequency identification), while only be allowed to make entries in a dedicated area of the social focus data section (Fig. 2). The write permissions include data integration from fitness watches, training performance in rehab or sports facilities and or self-collected nutritional data while interfaces to e.g. cell phone apps are available to substitute manual input. However, the medical staff, on the other hand, can read medically relevant data and has the right to make entries in the medical focus data (Fig. 2) showing who made the entry. This corresponds to an entry in the EHR stored in HIS (Fig. 7). Producers of medical products only have access to the medical device data area summarized as production (Fig. 7). Technical focus data (Fig. 2), e.g. appropriate revision instruments (c.f. chapter 3.5), are stored here while the patient and implant are linked in the medical focus data base (Fig. 2 and Fig. 7). This information is requested from the manufacturer via a modified procurement process, which ensures that the agreed data is available before the invoice for the implant is paid. Consequently, all relevant product information is stored enabling to simplify follow-up treatments, support minimally invasive interventions, and/or exclude medical interaction. The information transfer towards the manufacturer (c.f. chapter 3.3), on the other hand, can be enabled using a data integration centre with a data use and access committee for research inquiries as is currently being developed by Prokosch et al. [47]. The last role in the person category is the first responder (Fig. 8) which is introduced to explain the dynamic access approach. The first responder occurs in case of an accident or emergency if e.g. life-saving measures are necessary. For example, access is granted for a certain period if several predefined factors that were detected using a fitness watch or other smart device, take effect at the same time (e.g. oxygen saturation in the blood, blood pressure, and/or other health-endangering characteristics). However, these attributes are securely transferred to the LAB 4.0 database management system (chapter 4.1) to obtain the necessary information depending on the authorization or to allow to add data. For this purpose, a standardized, well-defined interface is used to realize the data exchange and integrate smart devices for pure information retrieval as well as to develop software extensions that can be used to store the data in the database while complying with access restrictions. This in order to create a digital ecosystem for different participants to provide patients with optimal and, above all, digital, end-to-end healthcare while providing adaptive access regulations meeting authenticity requirements while assuring authenticity appropriate access tracking.

*4.3. Data integration scenario*

Using the hospital database LMS 4.0 (Fig. 8) all individual elements of data ecosystem are presented and explained in reference to Fig. 10. In fact, the data ecosystem is divided in 4 levels: "Data storage", "Data harmonization", "Interfaces" and "Data input/Data output" are being distinguished for functional structuring. The "Data storage" level contains different relational databases, which, again, contain medically relevant data (medical device (chapter 3.3), patient record (Chapter 3.5) and health (Chapter 3.4). Data pre-processing is performed at the "Data harmonization" level (c.f. Fig. 8 and Fig. 10) which means, that incoming data is adapted to the requirements of the LAB 4.0 database structure (Fig. 8) and sorted while outgoing (anonymized) data (e.g. via the data integration centre) is transferred via defined data exchange procedures with strictly recorded accesses. In the underlying, but closely related, "Interfaces" level (Fig. 10), interfaces are established by extensible middleware to communicate with the "Hospital database" (c.f. Fig. 8). This enables to integrate data users from different domains and query data from the database. The three levels "Interfaces", "Data harmonization" and "Data storage" are subjected to CIA triad (Confidentiality, Integrity, Availability) and ensure the functionality of the system. The "Data input / Data output" level connects the "Hospital database" with the environment. For example, manufacturers of medical devices can transfer product-specific data into the database to make it available to hospital staff (c.f. Fig. 3 and Fig. 10). Likewise, patients can store their e.g. vital signs from wearables in this database to e.g. support long-term examinations or to enable access in medical emergencies through attributed-based access control (Chapter 4.2). At the same time, patients can see their ERH, read digital doctor's notes, or view exam results. Physicians have interfaces to both connect medical exam machines to the database and write data, as well as store information by medical staff while the data can be viewed hospital-wide and processed with appropriate IT systems.

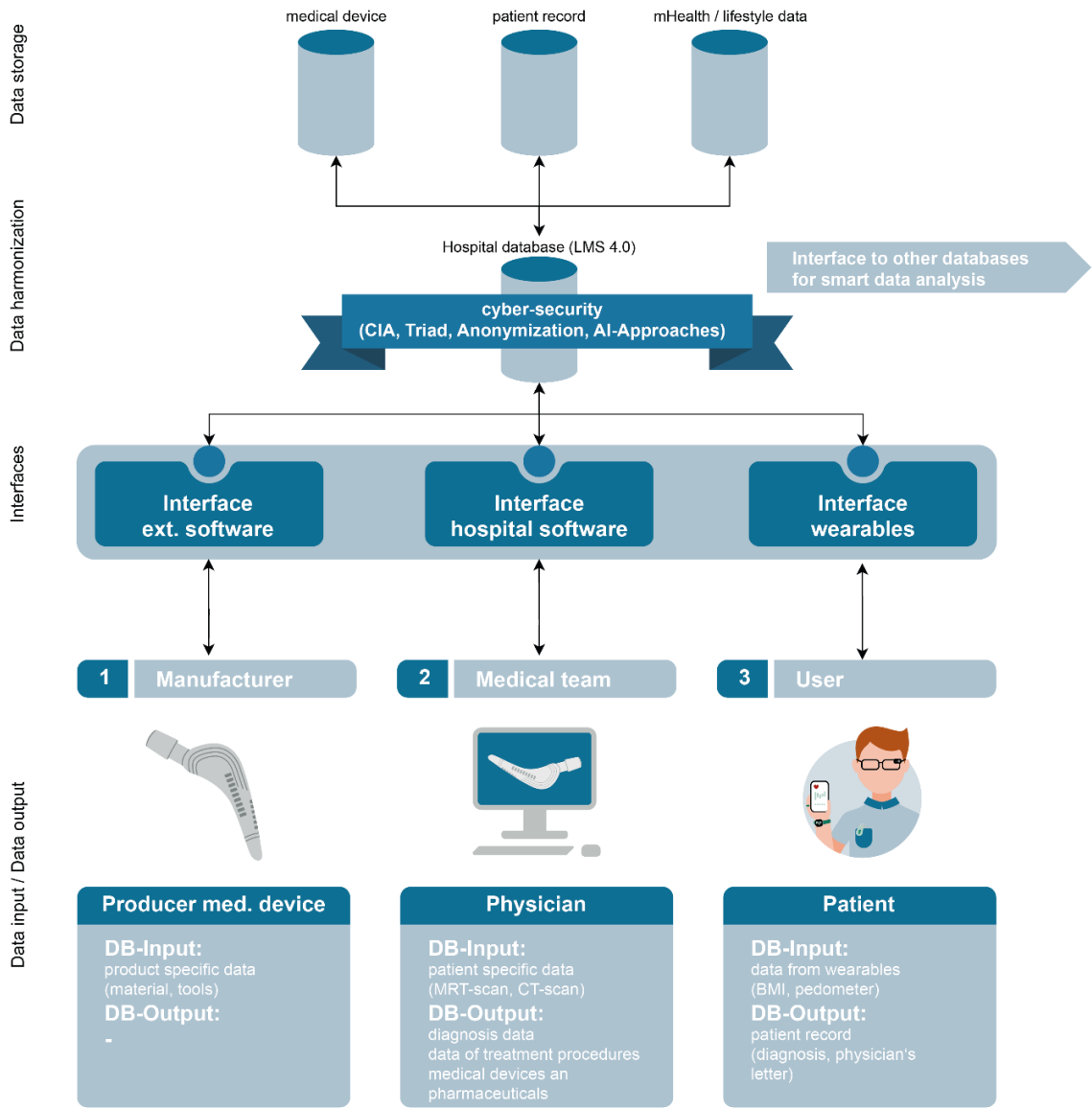

**Figure 10.** Data integration ecosystem with hierarchical data access.

## 5. Discussion

The overriding goal is to use digitization to improve the networking of interdisciplinary domains and to create secure interfaces for exchange as a prerequisite for intelligent data analysis. For this purpose, a representative hip implant application scenario was chosen due to an existing network of social, medical and technical domains. Moreover, the interaction resulting from a skill- and experience-based union of implant and recipient results in an individual constellation that is subject to change over time. A key element in the resulting constellation is the UDI based on an inherent feature, which can be read out non-invasively after implantation. The permanently readable feature acts as a key to technical focus data which represents testable and / or documented properties that are made available by the provider and which are of direct or downstream interest. Consequently, it was shown how the technical focus data can be integrated into existing data ecosystems. This, however, was only approached at the hospital level, which is explained by the unclear legal framework and the missing data infrastructure for a broader context. Nevertheless, the consolidation of distributed databases in a single technology solution is a scalable concept that can be transferred from a single hospital to a global solution. Another important aspect is the introduced hierarchical data access based that is based on a shell embedded role model and staggered user rights. Here, the attributed-based access control shall be emphasized because this represents non-rigid boundary conditions in preparation for future regulations. The selected user profiles and the granted rights, on the other hand, are only examples that are up for discussion and which need to be specified and challenged in further research. However, the data integration scenario distinguishes four levels of action layering data storage, data harmonization, interfaces and the data input/data output layer, which harmonizes the application scenario and the digital ecosystem. Nevertheless, future research must show how real benefit can be created by data linkage and how this can be monetized. Balancing the personal rights of the individual while achieving sustainable technological innovation is seen as the central challenge, which must be faced in a global context.

## 6. Conclusions

Personalized medicine requires cross-domain linkage of data which, in turn, requires an appropriate data infrastructure and adequate hierarchical data access solutions,

The hip implant is a prime example of the usefulness of cross-domain linkage of data because it bundles social factors of the individual patient, medical aspects in the context of the implantation and technical aspects of the implant,

Unique Device Identification in terms of inherent identifiers can be the key to (selective) long-term data access especially if the postoperative readout is guaranteed,

Selective laser melting and/or electron beam melting offer the possibility to integrate inherent features already in the design process what enables to close the traceability gap,

It is necessary to open existing databases using suitable interfaces for secure integration of data from end devices (e.g. wearables and/or end users) and to assure availability through suitable access models (role-based, attribute-based, hybrid) while enabling to guarantee long-term independent data persistence,

A suitable strategy requires the combination of technical solutions from the areas of data storage, cryptographic procedures and software engineering as well as organizational changes among the actors involved (e.g. hospital staff, implant manufacturers, patients),

Holistic approaches require interdisciplinary cooperation and cross-domain data spaces, while innovative approaches and services must be developed prior or parallel to the ongoing clarification of the legal framework conditions,

In order to provide viable and transferable solutions at the time of legal clarification cross-domain lighthouse projects are needed in order to assure the timely availability of digital business models, suitable data alliances and an adequate digital infrastructure.

**Author Contributions:** Conceptualization, K.K., A.S. and U.T.; methodology, A.S.; software, K.K. and G.L.; investigation, K.K., N.M, C.N. and G.L; resources, S.I. and W.G.D.; writing—original draft preparation, A.S.; writing—review and editing, K.K., A.S., N.M., C.N., U.T., G.L., A.B.A., S.I. and W.G.D.; visualization, A.S., U.T., K.K., N.M, C.N., A.B.A., and G.L.; supervision, A.S. and S.I..; project administration, K.K., A.S. funding acquisition, S.I. and W.G.D. All authors have read and agreed to the published version of the manuscript.

**Funding:** presented "Unique device identification based linkage of hierarchically accessible data domains in prospective hospital data ecosystems" is part of the Fraunhofer Light-house Project "futureAM - Next Generation Additive Manufacturing" funded internally by the Fraunhofer-Gesellschaft e.V.

**Conflicts of Interest:** "The authors declare no conflict of interest."